\documentclass[12pt]{article}
\usepackage[T1]{fontenc}
\usepackage[all]{xy}
\usepackage{float}
\usepackage{amsmath}
\usepackage{graphicx}
\usepackage[super,sort&compress]{natbib}
 
\newcommand{\phiEq}{\phi_\mathrm{eq}}
\title{Theory of droplet ripening in stiffness gradients}
\author{Estefania Vidal-Henriquez$^1$ and David Zwicker$^1$}%
\date{%
	$^1$~Max-Planck Institute for Dynamics and Self-Organization, Am Fa\ss{}berg 17, 37077 G\"{o}ttingen, Germany.
}
\begin{document}
\maketitle

\begin{abstract}
	Liquid-liquid phase separation is an important mechanism for compartmentalizing the cell's cytoplasm, allowing the dynamic organization of the components necessary for survival.
	However, it is not clear how phase separation is affected by the complex viscoelastic environment inside the cell.
	Here, we study theoretically how stiffness gradients influence droplet growth and arrangement.
	We show that stiffness gradients imply concentration gradients in the dilute phase, which transport droplet material from stiff to soft regions.
	Consequently, droplets dissolve in the stiff region, creating a dissolution front.
	Using a mean-field theory, we predict that the front emerges where the curvature of the elasticity profile is large and that it propagates diffusively.
	This elastic ripening can occur at much faster rates than classical Ostwald ripening, thus driving the dynamics.
	Our work shows how gradients in elastic properties control the size and arrangement of droplets, which has potential applications in soft matter physics and plays a role inside biological cells.
\end{abstract}

\section{Introduction}

Phase separation has recently been established as a crucial mechanism for organizing membrane-less organelles in biological cells.\cite{brangwynne2009germline, Berry2018, Boeynaems2018a, Alberti2019a}
However, these membrane-less organelles often posses properties of liquid-like droplets, like internal rearrangement, spherical shapes, and monomer exchange with their surrounding.
These droplets exist in complex environments that cannot be described as simple liquids.
For example, the cytoskeleton in the cytosol~\cite{Pegoraro2017,Gardel2008} as well as the chromatin in the nucleoplasm~\cite{erdel2015viscoelastic} have solid-like properties, which could affect droplets.
Indeed, recent experiments showed that droplets typically form in chromatin-poor regions in the nucleus.~\cite{shin2018liquid}
This suggests that the elastic properties of chromatin suppress droplet formation.
However, it is difficult to disentangle the effect of the elastic surrounding from other potential processes that could affect droplet formation in the complex case of a living cell.


Physical experiments with synthetic materials can
help to understand how an elastic environment affects droplet growth.
For instance, oil droplets growing in a homogeneous PDMS gel form mono-disperse emulsions and the stiffness of the gel controls droplet size.~\cite{style2018liquid}
There are several advantages of these experiments:
First, the system is not driven, implying it relaxes toward equilibrium after preparation.
Second, the gel is strongly cross-linked, so it does not spontaneously rearrange on the time scale of the experiments.
Viscous relaxation is thus negligible.
Third, droplets are large compared to the mesh size, implying that the gel can be described by a continuum theory.
Taken together, these properties allow to isolate the effects of an elastic environment on droplets.

The aim of the present paper is to understand how spatially varying stiffnesses affect droplet dynamics.
This is motivated by recent experiments showing that stiffness gradients lead to \textit{elastic ripening}, where droplets dissolve in stiffer regions.~\cite{rosowski2019elastic}
Moreover, these experiments revealed a dissolution front invading stiffer regions, while the material of the dissolving droplets accumulated in softer regions.
We developed a theoretical description of the situation, which is based on the assumption that the gel exerts a pressure onto droplets that is proportional to the local stiffness.~\cite{rosowski2019elastic}
Numerical simulations of this theory showed excellent agreement with the measured data.
Here, we analyze this model in detail and derive a simplified, coarse-grained version that allows us to predict where and when the dissolution front starts and how it evolves in time.
\section{Elasticity gradients produce dissolution fronts}



We aim at understanding the dynamics of an emulsion embedded in a gel with spatially varying stiffness.
Motivated by elastic ripening experiments~\cite{rosowski2019elastic}, we focus on the case where the gel is strongly cross-linked and behaves as an elastic material.
Moreover, the droplets are well separated and deform the gel only locally, so they only interact by exchanging material via the dilute phase.
This exchange is driven by the difference between the concentration in the dilute phase and the concentration right outside a droplet's surface.
We determine the latter by considering a single droplet in an homogeneous elastic environment.


\subsection{Thermodynamics of isolated droplets}

Motivated by recent experimental results \cite{rosowski2019elastic}, we describe a three-component system of two liquids and a gel.
When the temperature is lowered, one liquid forms droplets by segregating from the other two components.
Therefore, this phase separation can effectively be described as a binary system; droplets with a high volume fraction~$\phi_\mathrm{in}$ of the segregating liquid coexist with a dilute phase of lower volume fraction~$\phi$.
In the case of thermodynamically large phases, these equilibrium volume fractions can be determined by minimizing the total free energy of the system using the Maxwell construction~\cite{weber2019physics}; see SI.
In particular, the equilibrium conditions imply that the pressures and chemical potentials are identical in both phases.

In the case of a single spherical droplet of radius~$R$ embedded in an isotropic elastic matrix, the droplet exhibits an additional pressure $P$ due to both surface tension and elastic effects.
In the simple case of a dense droplet phase and an ideal dilute phase, the resulting equilibrium volume fraction in the dilute phase can be approximated as
\begin{equation}
\phi_\mathrm{eq} \approx \phi_0 \exp\left(\frac{P}{c_\mathrm{in} \, k_\mathrm{B} T}\right)
\;;
\label{Eq_ApproxPhiEq}
\end{equation}
see SI.
Here, $c_\mathrm{in}$ is the concentration of the segregating fluid inside the droplet, $k_\mathrm{B}$ is Boltzmann's constant, $T$ is absolute temperature, and
$\phi_0$ denotes the equilibrium volume fraction of the dilute phase in a thermodynamically large system for $P=0$. 
Consequently, Eq.~\eqref{Eq_ApproxPhiEq} implies that the additional pressure~$P$ increases the equilibrium volume fraction in the dilute phase.


The pressure~$P$ on a single spherical droplet of radius~$R$ embedded in an elastic gel is 
\begin{equation}
P = \frac{2\gamma}{R} + P_E(R)
\label{eqn:droplet_pressure}
\;,
\end{equation}
where $\gamma$ is the surface tension and $P_E$ is the pressure exerted by the isotropic elastic gel, which is related to its stress-strain curve.
We here consider droplets that are much larger than the gel's mesh size, implying that the gel experienced large strains during droplet growth.
Such situations typically arise when the gel has a maximal stress~$P_\mathrm{C}$ that it can sustain. 
For example, when growing droplets fracture the gel, $P_\mathrm{C}$ is the critical stress that is required for fracturing.
In the simplest case, $P_\mathrm{C}$ is proportional to the Young's modulus~$E$ of the gel, $P_\mathrm{C} = \zeta \, E$.
The proportionality constant~$\zeta$ can be determined experimentally or from theory.
For example, Neo-Hookean theory\cite{rivlin1948hydrodynamics,rivlin1951large} predicts $\zeta=5/6$, while the silicon gels used in the elastic ripening experiments\cite{kim2018extreme} exhibit $\zeta\approx0.5$.
Taken together, we here focus on the case where the pressure exerted by the gel is a simple function of the stiffness, $P_E=\zeta E$.

To see whether surface tension~$\gamma$ is important in the ripening experiments, we next estimate the relevant pressure gradients.
The pressure gradient created by surface tension is roughly $\gamma/(R\ell)$, where $\gamma/R$ is an estimate for the pressure difference between droplets and $\ell$ is their typical separation.
Conversely, the pressure difference created by the external gel can be estimated as $E/w$, where $E$ is a typical stiffness and $w$ is the length scale over which it varies.
In the ripening experiments, we find $\gamma/(R\ell) \ll E/w$, implying that surface tension is negligible.\cite{rosowski2019elastic}
Taken together, the equilibrium volume fraction $\phiEq$ can thus be approximated by
\begin{equation}
\phi_\mathrm{eq}(E) \approx \phi_0 \exp\left(
\frac{ E}{\hat{E}}
\right)
\;,
\label{eqn:phi_eq}
\end{equation}
where $\hat{E}=c_\mathrm{in} \, k_\mathrm{B} T/\zeta$ is the relevant stiffness scale.
This expression allows us to determine the volume fraction~$\phi_\mathrm{eq}$ outside a droplet embedded in a gel described by a local stiffness~$E$.


\subsection{Dynamical equations of emulsions}
We now consider an emulsion of droplets embedded in a gel with spatially varying stiffness~$E(\vec{x})$.
We describe the system by the droplet radii $R_i$ and their positions $\vec{x_i}$, as well as the volume fraction~$\phi(\vec{x})$ in the dilute phase.
The thermodynamics discussed in the previous section imply that the equilibrium volume fraction~$\phi_\mathrm{eq}$ right outside each droplet depends on its position.
The difference between $\phi_\mathrm{eq}$ and $\phi$ drives a diffusive flux between the droplet and the dilute phase.
Since we are only interested in dynamics on length scales larger than the droplet radii, we evaluate all involved quantities at the droplet position~$\vec{x_i}$.
Consequently, the material efflux~$J_i$ integrated over the droplet surface can be expressed as $J_i=4\pi D R_i [\phi_\mathrm{eq}(\vec{x_i}) - \phi(\vec{x_i})]$, where $D$ is the molecular diffusivity.\cite{weber2019physics}
This flux drives changes in droplet radius,
\begin{equation}
\dfrac{\mathrm{d}R_i}{\mathrm{d}t} = \dfrac{D}{R_i\phi_\mathrm{in}}\left[
\phi(\vec{x_i})-\phi_\text{eq}(\vec{x_i})
\right]
\;,
\label{eqn:full_model_R}
\end{equation}
where we used that the volume fraction~$\phi_\mathrm{in}$ inside the droplet is much larger than the fraction~$\phi_\mathrm{eq}$ outside.
This equation describes how a droplet grows by taking up material from its immediate surrounding.
On large length scales, material diffuses in the dilute phase, implying
\begin{equation}
\partial_t \phi = D \nabla^2 \phi - \phi_\mathrm{in}\sum_i \dfrac{\mathrm{d}V_i}{\mathrm{d}t} \delta(\vec{x_i} - \vec{x})
\;,
\label{eqn:full_model_phi}
\end{equation}
where $V_i=(4\pi/3) R_i^3$ are the individual droplet volumes.
Here, the last term accounts for the material exchange with droplets.
For simplicity, we consider immobilized droplets whose positions~$\vec{x}_i$ are constant.
Consequently, Eqs. \ref{eqn:full_model_R}--\ref{eqn:full_model_phi} fully describe how an emulsion of droplets evolves over time.

The dynamics of the system is governed by two diffusive fluxes that act on different length scales.
Locally, material is exchanged between the droplets and the dilute phase by the flux~$J_i$.
Conversely, transport on longer length scales only happens in the dilute phase by the redistribution flux $-D\nabla \phi$.

\subsection{Numerical simulations show dissolution fronts}

\begin{figure}[h!]
	\centering
	\includegraphics[width=0.7\linewidth]{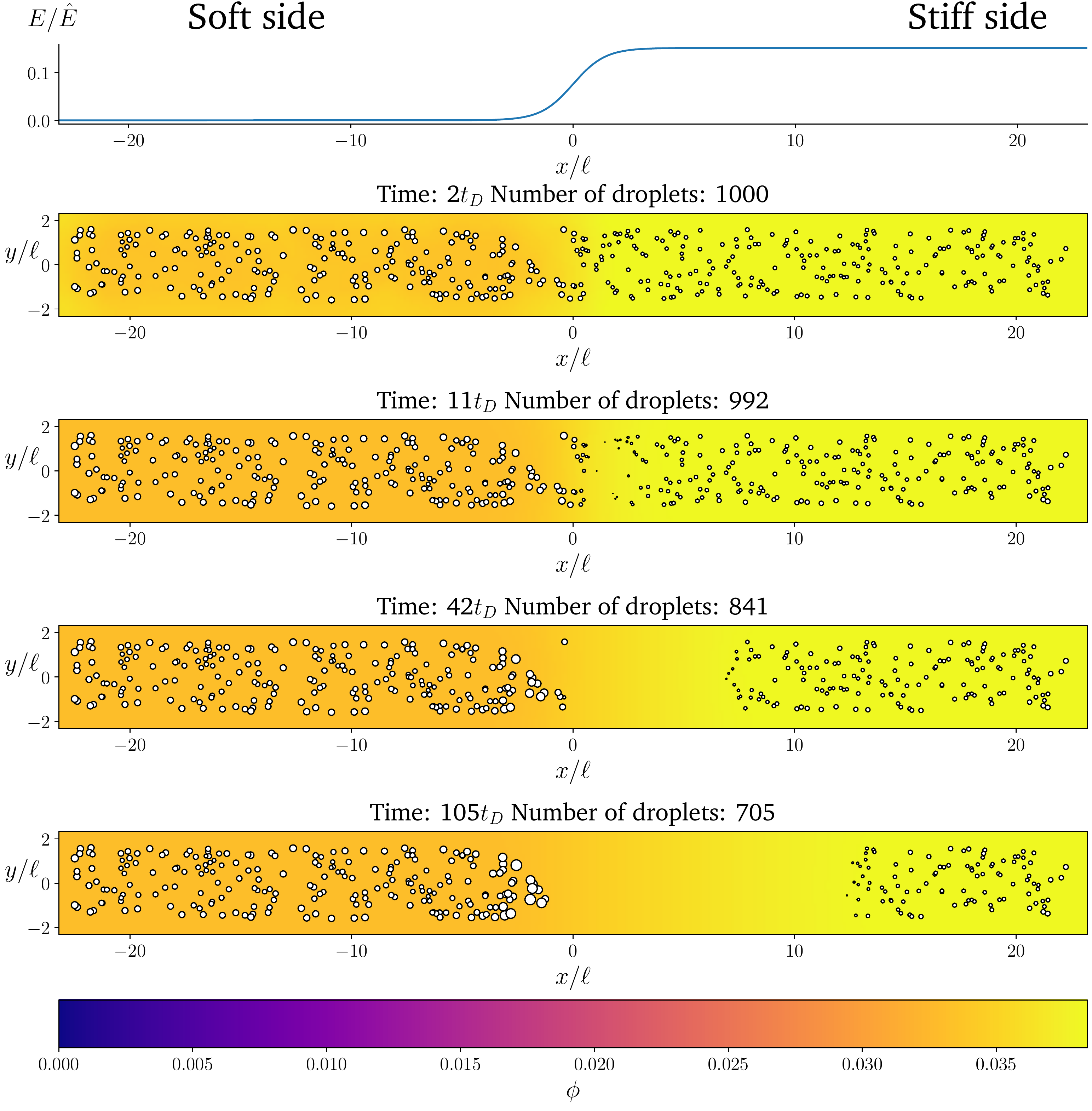}
	\caption{Numerical simulation showing a dissolution front invading the stiff region defined by a sigmoidal elasticity profile (upper panel).
		Subsequent images show simulation snapshots (obtained by solving Eqs. \ref{eqn:full_model_R}--\ref{eqn:full_model_phi}) of the droplets (symbols) and the volume fraction~$\phi$ in the dilute phase (density plot with color bar at the bottom) at the indicated times.
		Model parameters are $\phi_0=0.033$, $\phi_\mathrm{in}=1$, $\psi_\mathrm{stiff}=0.09\phi_0$, $E_\mathrm{stiff}=0.15\,\hat{E}$, $E_\mathrm{soft}=10^{-4}\,\hat{E}$, and $w = 1.45 \, \ell$.
		Here, $\ell=(V_\mathrm{sys}/N_\mathrm{drop})^{1/3}$ is a typical droplet separation with associated diffusive time $t_\mathrm{D} = \ell^2/D$, where $V_\mathrm{sys}$ is the system's volume and $N_\mathrm{drop}$ is the total number of droplets.
	}
	\label{Figure_background}
\end{figure}

We simulated Eqs. \ref{eqn:full_model_R}--\ref{eqn:full_model_phi} to understand the effects of an elasticity gradient on the emulsion dynamics.
Motivated by elastic ripening experiments~\cite{rosowski2019elastic}, we consider a system consisting of two regions of different stiffness $E_\mathrm{soft}$ and $E_\mathrm{stiff}$.
When the two regions are put side-by-side, a transition region emerges.
To capture this, we model the stiffness profile in the entire system by
\begin{equation}
E(x)=
\frac{E_\text{stiff} + E_\text{soft}}{2} + 
\dfrac{E_{\text{stiff}}-E_{\text{soft}}}{2}
\tanh\left(\frac{x}{w}\right)
\;,
\label{eqn:stiffness_profile}
\end{equation}
where $x$ denotes the coordinate perpendicular to the interface and $w$ is the width of the transition region.
Generally, we chose parameter values motivated by experiments~\cite{rosowski2019elastic}.
For instance, we initialized the simulations with tiny droplets distributed uniformly in the whole system and we imposed a uniform volume fraction field $\phi(\vec x)$ in the dilute phase.

Fig.~\ref{Figure_background} shows the time course of a typical simulation.
Starting from the homogeneous initial condition, the system quickly forms two separate regions aligned with the stiffness profile (upper panel).
Here, the stiff side exhibits smaller droplets and larger volume fractions in the dilute phase compared to the soft side.
Droplets then start dissolving in the transition region and a dissolution front moves into the stiff side.
Simultaneously, droplets grow on the soft side of the transition region while droplets far into the soft side remain unchanged.
These dynamics can be understood qualitatively by considering the diffusive fluxes in the system.


In the initial stage, the system is supersaturated everywhere, $\phi > \phi_\mathrm{eq}$.
Consequently, material is transferred from the dilute phase to the droplets until a local equilibrium is reached, $\phi = \phi_\mathrm{eq}$.
Eq.~\eqref{eqn:phi_eq} implies that $\phi_\mathrm{eq}$ is smaller for softer regions, so more material is absorbed by the droplets.
We thus observe larger droplets in softer regions (see Fig.~\ref{Figure_background}), consistent with experimental observations~\cite{style2018liquid}.

After the initial, local equilibration, material redistribution on longer length scales sets in.
Since the stiffer side exhibits larger volume fractions~$\phi$ in the dilute phase, material is transported to the soft side.
Consequently, on the stiff side, $\phi$ drops below the local equilibrium volume fraction~$\phi_\mathrm{eq}$, droplets shrink, and eventually dissolve.
This process starts close to the transition region, since the redistribution flux is driven by gradients in $\phi$, which do not exist further away.
Once droplets start disappearing in the transition region, droplets further away begin to be affected and a dissolution front forms that invades the stiff side.
All the material redistributed from the stiff side is absorbed by the droplets on the soft side close to the transition region, which effectively shield all the other droplets on the soft side.

\section{A coarse-grained model explains the dissolution dynamics}

To understand the front's dynamics quantitatively, we next consider a simplified version of our model.
Here, we do not describe the dynamics of individual droplets, but rather focus on the fractions of material in droplets and in the dilute phase.
We thus introduce the coarse-grained volume fraction~$\psi$ of material contained in droplets,
\begin{equation}
\psi(\vec{x}, t) = \phi_\text{in}\dfrac{\iiint \sum_i V_i \delta(\vec{x_i}-\vec{y}) \, \mathrm{d}^3y}{\iiint \mathrm{d}^3y}
\;,
\end{equation}
where the integrals are performed over a small discretization volume centered at $\vec{x}$.
Note that the discretization volume needs to be large enough to contain multiple droplets, but also small compared to the characteristic length scales of the elasticity gradient.
Introducing the local mean droplet volume~$V(\vec{x}, t)$, we can express $\psi$ as
$\psi=\phi_\mathrm{in} n V$, where $n$ is the local droplet number density.
Motivated by the elastic ripening experiments and our numerical simulations, we consider the case where the volume of individual droplets does not deviate substantially from the local mean volume~$V$.
In this case, the volume fractions $\phi$ and $\psi$, characterizing the amount of material in the dilute phase and droplets, respectively, are sufficient to describe the system's state.

The dynamics of the coarse-grained system follow from Eqs. \ref{eqn:full_model_R}--\ref{eqn:full_model_phi}.
We show in the SI that the dynamical equations are
\begin{subequations}
	\begin{align}
	\partial_t\psi&=    
	D\left(\phi-\phi_\text{eq}\right)\left(\dfrac{48\pi^2n^2}{\phi_\text{in}}\psi\right)^{1/3}
	\\
	\partial_t\phi&=D\nabla^2\phi-\partial_t\psi
	\label{eqn:reduced_model_dilute}
	\;.
	\end{align}
	\label{Eq_Reduced_Model}
\end{subequations}
The first equation describes the local exchange of material between droplets and their surrounding, while the second equation accounts for the redistribution of material over long length-scales.

\begin{figure}[h!]
	\centering
	\includegraphics[width=0.75\linewidth]{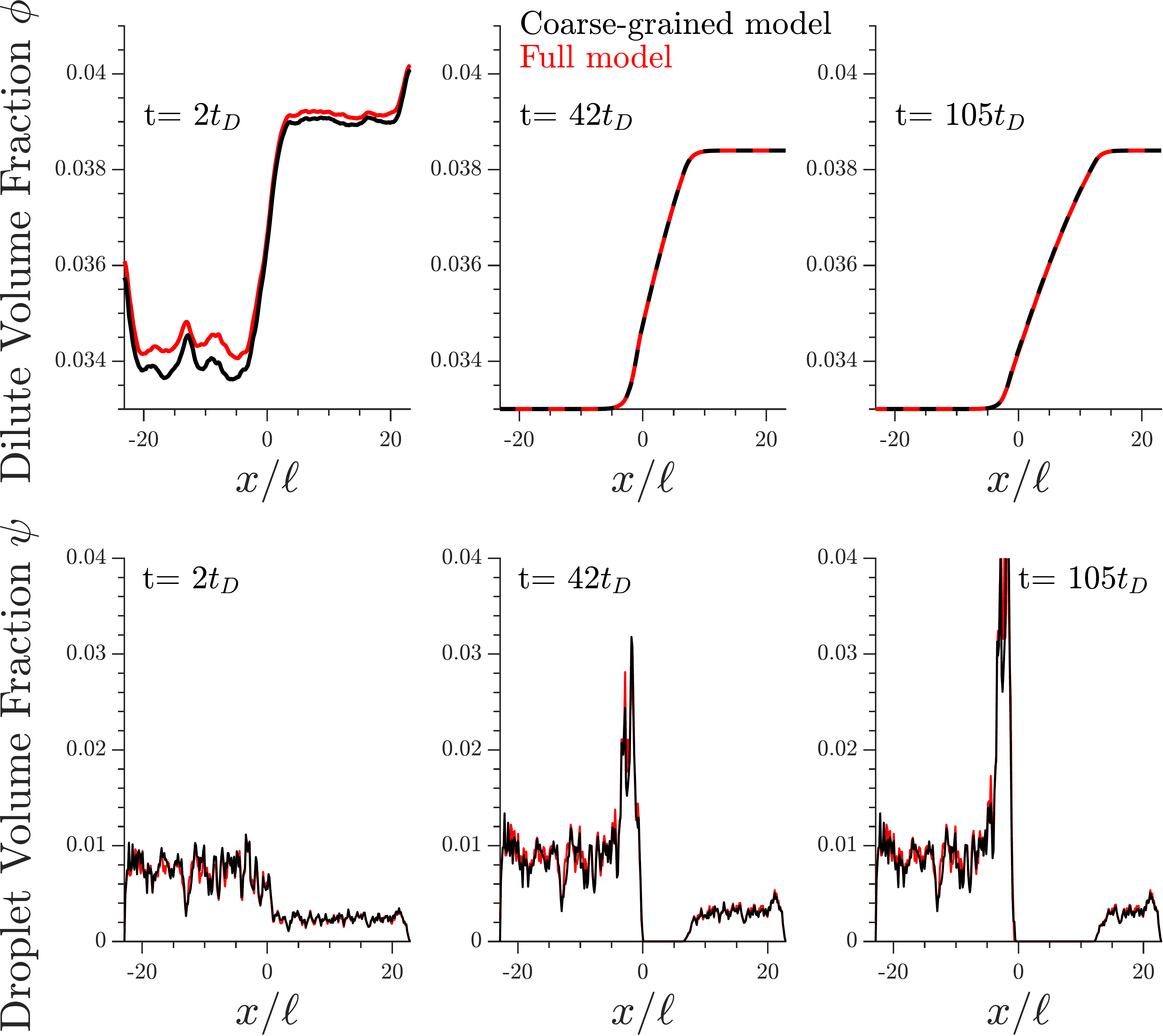}
	\caption{
		The coarse-grained model (black lines, Eq. \ref{Eq_Reduced_Model}) captures the detailed dynamics of the full model (red lines, Eqs. \ref{eqn:full_model_R}--\ref{eqn:full_model_phi}).
		Shown are the profiles of the volume fractions $\phi$ in the dilute phase (upper panels) and the fraction~$\psi$ contained in droplets (lower panels) for three different time points~$t$.
		Model parameters are identical to Fig.~\ref{Figure_background}.
	}
	\label{Figure_reduced_Comparison}
\end{figure}

Fig.~\ref{Figure_reduced_Comparison} shows that the results of the numerical simulation of the coarse-grained model are virtually indistinguishable from that of the detailed model.
Therefore, the coarse-grained model captures the essential features of the elastic ripening process.
In particular, the dynamics of the dissolution front are governed by the material distribution, while individual droplets are irrelevant.

\subsection{Dissolution starts at high curvatures of the elasticity profile}

We now use the coarse-grained model to understand where and when the dissolution front appears, i.e., when droplets first disappear.
In the numerical simulations, we observe that $\phi$ quickly approaches $\phiEq$ by local equilibration between the droplets and the dilute phase.
Assuming that the system is initialized with $\phi=\phi_\mathrm{init}$ and $\psi=\psi_\mathrm{init}$, the volume fractions approach $\phi\approx\phiEq$ and $\psi = \psi_0$ with $\psi_0 \approx \psi_\mathrm{init} + \phi_\mathrm{init} - \phiEq$ after this initial stage.
Consequently, the volume fraction~$\phi$ in the dilute phase is controlled by the stiffness profile $E(\vec x)$, while the remaining material is absorbed in the droplets.

After the initial equilibration stage, the inhomogeneities in $E$, and thus in $\phi$, drive diffusive fluxes toward the soft side.
However, we observe that these fluxes mostly affect $\psi$ and hardly change $\phi$ before the first droplets disappear.
To understand the dynamics in this stage, we approximate
Eq.~\eqref{eqn:reduced_model_dilute} by $\partial_t \psi \approx D\nabla^2 \phiEq$.
Consequently, $\psi$ evolves as 
\begin{equation}
\psi(\vec x, t) \approx \psi_0(\vec x) + t D \nabla^2\phiEq
\label{eqn:linear_droplet_dynamics}
\;.
\end{equation}
Consequently, the curvature of the equilibrium field~$\phiEq$, set by the elasticity profile, controls droplet dynamics.
In particular, droplets grow in convex regions ($\nabla^2\phiEq > 0$), while they shrink in concave ones.
Note that Eq.~\eqref{eqn:linear_droplet_dynamics} only holds when $\psi > 0$, since otherwise droplets would be absent and the flux in the dilute phase changes $\phi$; see Eq.~\eqref{eqn:reduced_model_dilute}.

We can use Eq.~\eqref{eqn:linear_droplet_dynamics} to estimate the time and position of the start of the dissolution front.
In particular, droplets dissolve after a time $\tau_*(\vec x) \approx -\psi_0(\vec x)/(D\nabla^2\phiEq)$, when all material is removed from the droplet phase.
The dissolution front starts at the earliest of these time points, $\tau_\mathrm{start}=\min_{\vec x}(\tau_*|\tau_* \ge 0)$, which is given by
\begin{equation}
\label{Eq_Time_approx}
\tau_\mathrm{start} = \frac1D\min_{\vec{x}}\left(
-\dfrac{\psi_0(\vec x)}{\nabla^2\phiEq}
\right)
\end{equation}
where the minimum is constrained to regions where $\tau_* \ge 0$, i.e., where droplets shrink ($\nabla^2\phiEq < 0$).
The location corresponding to the minimum denotes the starting position of the front.
Eq.~\eqref{Eq_Time_approx} highlights that the front appears where droplets are small and sparse (small $\psi_0$) as well as dissolve quickly (strongly negative curvature $\nabla^2\phiEq$).

The starting time~$\tau_\mathrm{start}$ can be estimated for the simple stiffness profile given by Eq.~\eqref{eqn:stiffness_profile}.
In particular, the droplet volume fraction~$\psi_0$ will be close to the value~$\psi_\mathrm{stiff}$ deep into the stiff side;
the curvature is approximately~$\nabla^2\phiEq\sim\Delta\phi /w^2$, where $\Delta\phi=\phiEq(E_\mathrm{stiff}) - \phiEq(E_\mathrm{soft})$ denotes the difference in the equilibrium volume fractions between the two sides and $w$ is the width of the transition region.
Using these estimates, Eq.~\eqref{Eq_Time_approx} suggests a time scale
\begin{align}
\hat\tau &= \frac{w^2\psi_\mathrm{stiff}}{D \Delta\phi}
\label{eqn:starting_time_scale}
\;,
\end{align}
which should govern the starting time of the front.
In contrast, the starting position~$s_\mathrm{start}$ is dominated by the location of the largest negative curvature of~$\phiEq$.
For the simple stiffness profile given in Eq.~\eqref{eqn:stiffness_profile}, this position should scale with the width~$w$ of the transition region.

\begin{figure}[h!]
	\centering
	\includegraphics[width=0.75\linewidth]{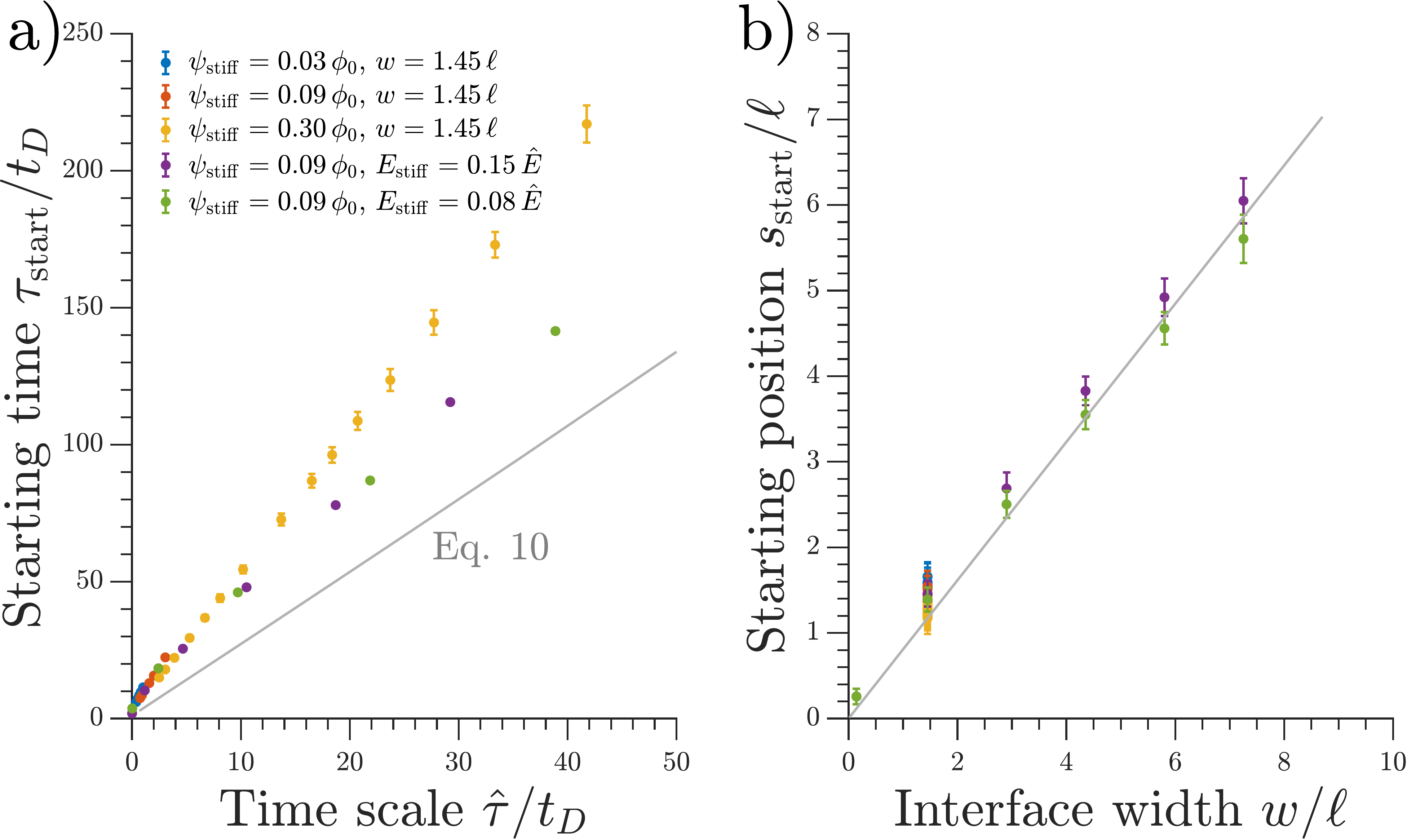}
	\caption{
		The starting time~$\tau_\mathrm{start}$ and position~$s_\mathrm{start}$ of the front scale with the predicted time and length scales, respectively.
		Results from numerical simulations of the full model (dots) for various parameters are compared to the theoretical predictions from the coarse-grained model (gray line).
		The time point $\tau_\mathrm{start}$ of the first dissolving droplet is shown in panel a) as a function of the predicted associated time scale $\hat\tau$ given in Eq. \eqref{eqn:starting_time_scale}.
		The theoretical prediction given by Eq. \eqref{Eq_Time_approx} is shown for $\psi=0.09\,\phi_0$ and $w=1.45\,\ell$.
		Panel b) shows the associated starting position $s_\mathrm{start}$ together with the equivalent prediction following from Eq. \eqref{Eq_Time_approx}, which is shown for $\psi=0.09\,\phi_0$ and $E_\mathrm{stiff}=0.15\,\hat E$.
		The remaining parameters are $E_\mathrm{soft}=10^{-4}\hat E$ and $\phi_\mathrm{in}=1$.
	}
	\label{Figure_starting_time}
\end{figure}

We test the prediction of Eq. \eqref{Eq_Time_approx} and the scaling discussed above by comparing to numerical simulations of the full model; see Fig.~ \ref{Figure_starting_time}.
The collapse of the starting times shown in the left panel indicates that $\hat\tau$ is the relevant time scale for this process.
Moreover, the actual prediction $\tau_\mathrm{start}$ given in Eq. \eqref{Eq_Time_approx} is within a factor of two of the measured data.
This analysis shows that the front appears earlier for larger stiffness differences (larger $\Delta\phi$) between the two sides, for narrower transitions regions (smaller $w$), as well as when there is less material in the droplet phase (small $\psi_\mathrm{stiff}$).
Fig.~\ref{Figure_starting_time}b shows the corresponding starting positions, which clearly are determined by the width~$w$ of the transition region.
The data collapse indicates that neither the absolute stiffnesses nor the droplet size affects where the front appears.

\subsection{Two fronts move in opposite direction initially}
Shortly after the first droplets dissolved, the surrounding droplets continue to shrink and disappear.
Consequently, the region devoid of droplets expands in all directions.
The material of the shrinking droplets is transported toward the soft side by diffusive fluxes.
Initially, the material will accumulate where the equilibrium field~$\phiEq$ has the largest positive curvature (maximal $\nabla^2\phiEq$); see Eq.~\eqref{eqn:linear_droplet_dynamics}.
The accumulating material is absorbed by the droplets in this region, which thus grow; see Fig.~\ref{Figure_background}.
The fact that droplets grow on the soft side implies that the front moving from $s_\mathrm{start}$ toward this side slows down.
Conversely, the front moving in the opposite direction can continue invading the stiff side.

\subsection{Dissolution front moves diffusively at late times}
We next analyze the late time dynamics of the front invading the stiff side.
Specifically, we consider the case where the width~$L$ of the region devoid of droplets is large compared to the width~$w$ of the transition region ($L \gg w$).
At this stage, the front moving toward the soft side is virtually stationary.
We can thus understand the dynamics of the front invading the stiff side by analyzing the width~$L$ of the region devoid of droplets (where $\psi=0$).
The dynamics in this region are governed by a simple diffusion equation of the volume fraction~$\phi$ in the dilute phase; see Eq. \eqref{eqn:reduced_model_dilute}.
At the stiff side of this region, the dissolving droplets impose the equilibrium fraction~$\phiEq(E_\mathrm{stiff})$ as a boundary condition for the diffusion equation.
The corresponding boundary condition at the soft side can be approximated by $\phiEq(E_\mathrm{soft})$.
For simplicity, we focus on slow fronts where the diffusion equation is in a stationary state, so the fraction~$\phi$ in the region devoid of droplets is governed by
\begin{align}
\phi(x) &=\phiEq(E_\mathrm{soft})  + \Delta\phi\frac xL
& \text{for} &&
0 &\le x \le L
\;,
\label{eqn:stationary_diffusion}
\end{align}
where $\Delta\phi=\phiEq(E_\mathrm{stiff})-\phiEq(E_\mathrm{soft})$ and $x$ denotes the distance from the boundary on the soft side.
The dynamics of $L$ can be obtained by considering the change of the total amount of material on the stiff side ($x\ge0$).
Because of material conservation, this change is equal to diffusive flux at $x=0$, which can be determined from Eq.~\eqref{eqn:stationary_diffusion}.
We show in the SI, that this implies $\partial_t L = \alpha/(2L)$ with
\begin{equation}
\alpha = 4D\left(1 + \frac{2\psi_\mathrm{stiff}}{\Delta\phi}\right)^{-1}
\label{eqn:front_diffusivity}
\;,
\end{equation}
where $\psi_\mathrm{stiff}$ is the droplet volume fraction deep into the stiff side.
Consequently, the region devoid of droplets expands as
\begin{equation}
L(t) = \sqrt{\alpha(t-t_0)}
\label{eqn:diffusive_front}
\;,
\end{equation}
where $t_0$ is such that $L(t_0) = 0$.
This equation implies a diffusive motion of the front with diffusivity~$\alpha$.

\begin{figure}[h!]
	\centering
	\includegraphics[width=0.75\linewidth]{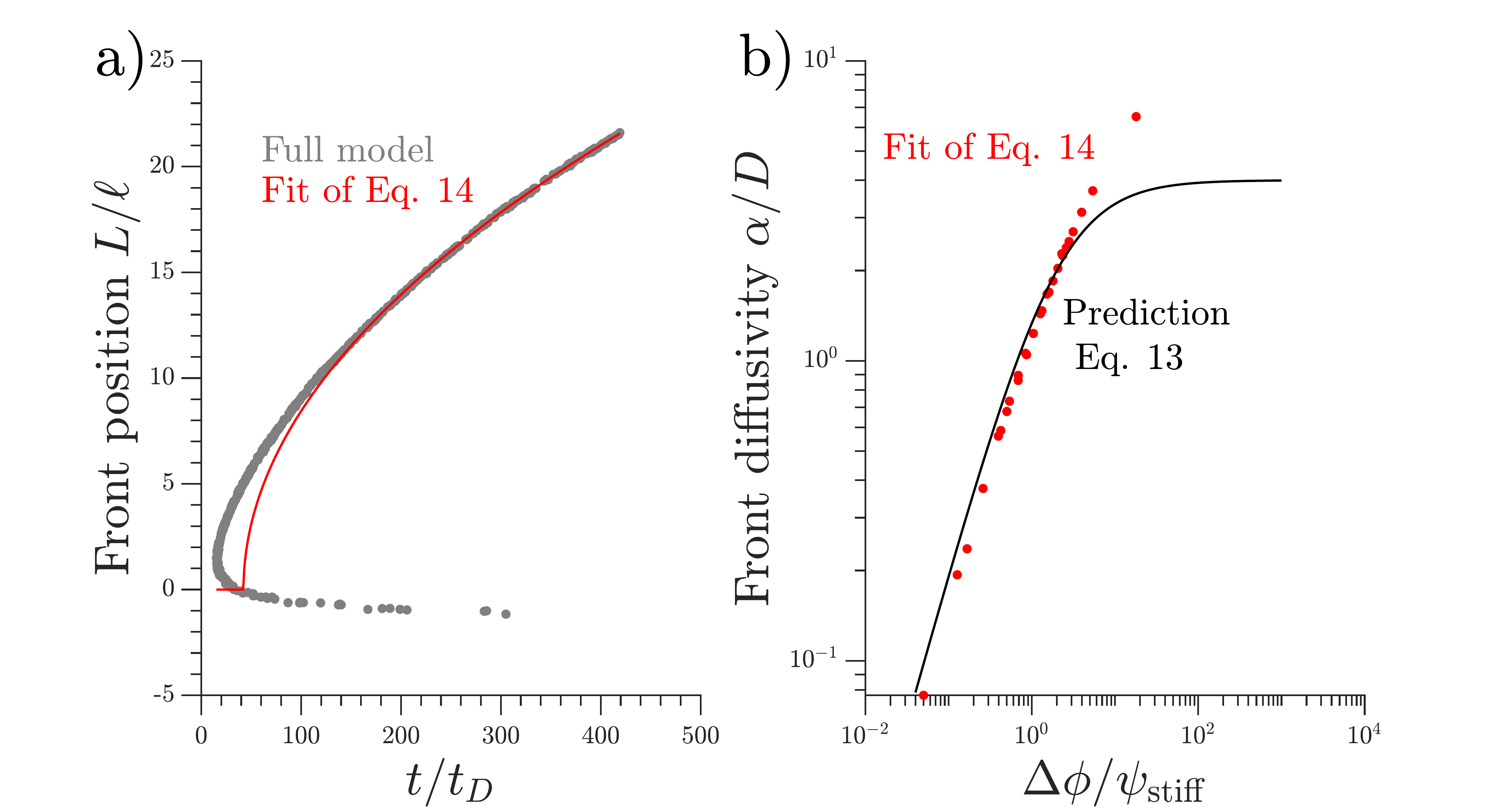}
	\caption{
		Dissolution fronts move diffusively.
		a) Position and time points of dissolving droplets in a simulation of the full model (dots) are compared to a fit (red line) of the theoretical prediction given in Eq. \eqref{eqn:diffusive_front}.
		Model parameters are $E_\mathrm{stiff} = 0.09\, \hat E$ and $\psi=0.09\,\phi_0$. Remaining parameters are given in Fig.~1.
		b) The front diffusivity~$\alpha$ (dots) determined from fitting to numerical simulations is compared to the prediction (line) given by Eq.~\eqref{eqn:front_diffusivity}.
		Simulations were done for $\psi_\mathrm{stiff}/\phi_0=0.03, 0.09, 0.30$ for various $E_\mathrm{stiff}$, while the remaining parameters are the same as in panel a).
	}
	\label{Figure_velocities}
\end{figure}

We test the theoretical prediction given in Eq.~\eqref{eqn:diffusive_front} by comparing to numerical simulations of the full model.
Fig.~\ref{Figure_velocities} shows the recorded times and positions when droplets dissolved (gray symbols), thus marking the dissolution fronts.
The fronts start in the transition region on the stiff side and then move in opposite directions.
The front moving toward the soft side slows down and comes to a halt on the soft side of the transition region, as predicted in the previous section.
Conversely, the front invading the stiff side is quicker and does not stop.
We measure its speed by fitting Eq.~\eqref{eqn:diffusive_front} to the front positions deep into the stiff side to extract $\alpha$ and $t_0$; see Fig.~\ref{Figure_velocities}a.
Since the model explains the measured data at late times, we conclude that the front moves diffusively.

The fitted front diffusivity $\alpha$ is presented in Fig.~\ref{Figure_velocities}b as a function of the relevant non-dimensional parameter $\Delta\phi/\psi_\mathrm{stiff}$.
This parameter compares the strength~$\Delta\phi$ of the elastic ripening to the fraction~$\psi_\mathrm{stiff}$ of material that needs to be removed from the droplets.
Consequently, the front is faster for larger $\Delta\phi/\psi_\mathrm{stiff}$.
The theoretical prediction for $\alpha$, given in Eq.~\eqref{eqn:front_diffusivity}, matches the data well for $\alpha < 2D$.
The fact that the front diffusivity~$\alpha$ needs to be smaller than or comparable to the molecular diffusivity~$D$ is not surprising since we assumed that the front is slow enough for the region devoid of droplets to be in a stationary state; see Eq.~\eqref{eqn:stationary_diffusion}.
Consequently, our theory predicts a maximal front diffusivity of $4D$ while the simulations indicate that faster fronts are possible.

\section{Elasticity profiles spatially control droplets}
\begin{figure*}[h!]
	\centering
	\includegraphics[width=0.9\linewidth]{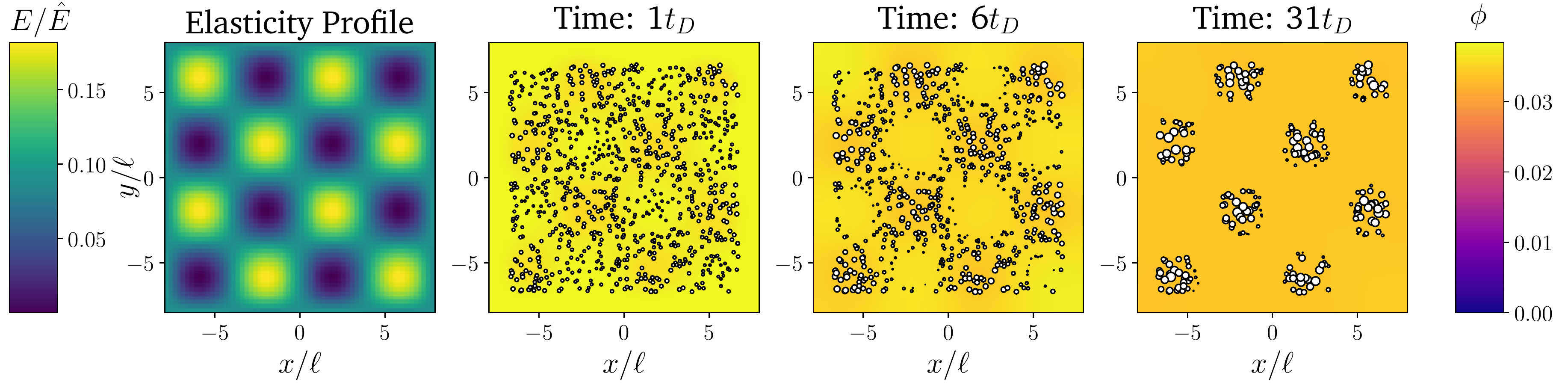}
	\caption{
		The elasticity profile controls the droplet dynamics.
		Left: Density plot of a two-dimensional elasticity profile~$E(\vec x)$.
		Right panels: Snapshots of a numerical simulation of the full model for three time points showing droplets as white symbols and the fraction~$\phi$ in the dilute phase as a color code.
		Remaining model parameters are $\phi_\mathrm{in}=1$,  $\phi_0=0.033$, and $\psi_\mathrm{stiff}=0.09\phi_0$.
	}
	\label{Figure_eggcarton}
\end{figure*}

So far, we considered elasticity profiles that only vary in one direction.
However, our full model (Eqs. \ref{eqn:full_model_R}--\ref{eqn:full_model_phi}) and the coarse-grained model (Eqs. \ref{Eq_Reduced_Model}) are more general.
In particular, Eq.~\eqref{Eq_Time_approx} implies that droplets first disappear in regions of strongly negative curvature $\nabla^2 \phiEq$.
Then, a front of dissolving droplets moves in all directions.
The material of the dissolving droplets accumulates in regions of large positive curvature $\nabla^2 \phiEq$, where droplets grow and remain for a long time.
Consequently, where droplets grow or shrink is governed by $\phiEq(\vec{x})$ and thus the elasticity profile $E(\vec{x})$.

Fig.~\ref{Figure_eggcarton} shows a numerical simulation of the full model for a two-dimensional elasticity profile (left panel).
A detailed simulation of a similar pattern has already identified that droplets accumulate in the soft valleys~\cite{shin2018liquid} and the time course shown in the right panels of Fig.~\ref{Figure_eggcarton} confirms that droplets follow the dynamics described above.
A movie of this simulation, as well as one for a more complex elasticity profile, can be found in the SI.
Taken together, this shows that we can engineer elasticity profiles to locate droplets in precise arrangements.

\section{Conclusions}
We presented a theoretical description of elastic ripening, which agrees quantitatively with experimental data.~\cite{rosowski2019elastic}
Therefore, our theory identifies the driving mechanism of elastic ripening: The elastic matrix exerts a pressure onto droplets that raises the equilibrium concentration in their surrounding;
Gradients in this concentration then lead to diffusive material transport in the system.
Surprisingly, the droplets do not start to dissolve in regions where the stiffness is maximal nor where its gradient is largest.
In fact, our coarse-grained model reveals that droplets initially shrink faster where the curvature of stiffness is larger.
However, at late times droplets only remain in the softest regions.
Taken together, we find complex dissolution dynamics, where for instance two fronts move in opposite directions, as in the experiments~\cite{rosowski2019elastic}.

The elastic ripening in stiffness gradients is similar to other droplet coarsening dynamics in gradient systems.
For instance, concentration gradients, e.g., of regulating species that compete for mRNA binding~\cite{Saha2016}, have been shown to bias droplet locations in experimental~\cite{brangwynne2009germline} and  theoretical studies~\cite{Lee2013,weber2017droplet,Kruger2018}.
Similarly, other external fields, like temperature gradients created by local heating~\cite{antonova2017local, choi2020probing} or even gravity~\cite{Feric2013} could be used to control droplet arrangements. 
Such systems can be analyzed using approaches that are similar to the ones presented here.

We showed that elastic ripening allows to control droplet arrangements, which could for instance be used in technical applications for micropatterning or for creating structural color.
Moreover, our theory can help to understand the localization of biomolecular condensates in biological cells.
For instance, elastic ripening explains experiments where droplets have been induced in the stiff regions of heterochromatin, but moved into softer regions immediately~\cite{shin2018liquid}.
We expect that similar processes happen in the cytosol, where biomolecular condensates should be less likely where the cytoskeleton is dense.
Interestingly, there are counterexamples, like centrosomes that localize to regions of high microtubule density~\cite{zwicker2014centrosomes,Redemann2017,Zwicker2018b} or ZO-1 clusters that concentrate in the acto-myosin cortex~\cite{Schwayer2019,Beutel2019}.
This seems to contradict elastic ripening, but in both examples the condensates interact with the elastic matrix: centrosomes bind the tubulin of microtubules~\cite{Baumgart2019} and the ZO-1 protein interacts with the F-actin of the cortex~\cite{Fanning2002}.
Consequently, there are two competing gradients in this situation: droplets are repelled by the stiffness of the surrounding matrix but are attracted by its molecular components.
Indeed, when the actin-binding domain of ZO-1 is removed, the clusters do not accumulate in the cortex anymore, but are more broadly distributed\cite{Schwayer2019}, as predicted by elastic ripening.

Our theoretical description of elastic ripening can be naturally extended to include other effects.
In fact, the dynamics described by Eqs.~\eqref{eqn:full_model_R}--\eqref{eqn:full_model_phi} already include surface tension effects when the pressure given by Eq.~\eqref{eqn:droplet_pressure} is used.
Although we did not analyze the impact of surface tension since it is negligible in the experiments, it will become important on longer timescales or when surface tensions are large.
Moreover, the elastic properties of the matrix might be more complex then considered here.
For instance, the cytoskeleton can shown strain-stiffening, which might arrest droplet growth, as well as visco-elastic effects, which allow relax elastic stresses.~\cite{Gardel2008}
The latter effect can be captured by the theory of viscoelastic phase separation, which affects the coarsening behavior.~\cite{Tanaka2000,Tanaka2005}
Finally, the droplets themselves can possess elastic properties.~\cite{Jawerth2018}
They sometimes even form solid-like aggregates~\cite{Patel2015} that potentially cause diseases~\cite{Alberti2019a}.
All these effects might crucial for understanding the behavior of biomolecular condensates in cells.

\section*{Conflicts of interest}
The authors declare no conflicts of interest.

\section*{Acknowledgements}

We thank Tal Cohen, Eric Dufresne, Frank Jülicher, Peter Olmsted, Kathryn A. Rosowski, and Pierre Ronceray for helpful discussions.

\bibliography{rsc} 
\bibliographystyle{ieeetr} 

\renewcommand\thefigure{S.\arabic{figure}}   
\renewcommand{\theequation}{S.\arabic{equation}}
\setcounter{equation}{0}
\setcounter{figure}{0}

\newpage
\section*{Supplementary Material}
\subsection*{Equilibrium concentrations in the presence of external pressure}
\begin{figure}[tb]
	\centering
	\includegraphics[width=0.5\linewidth]{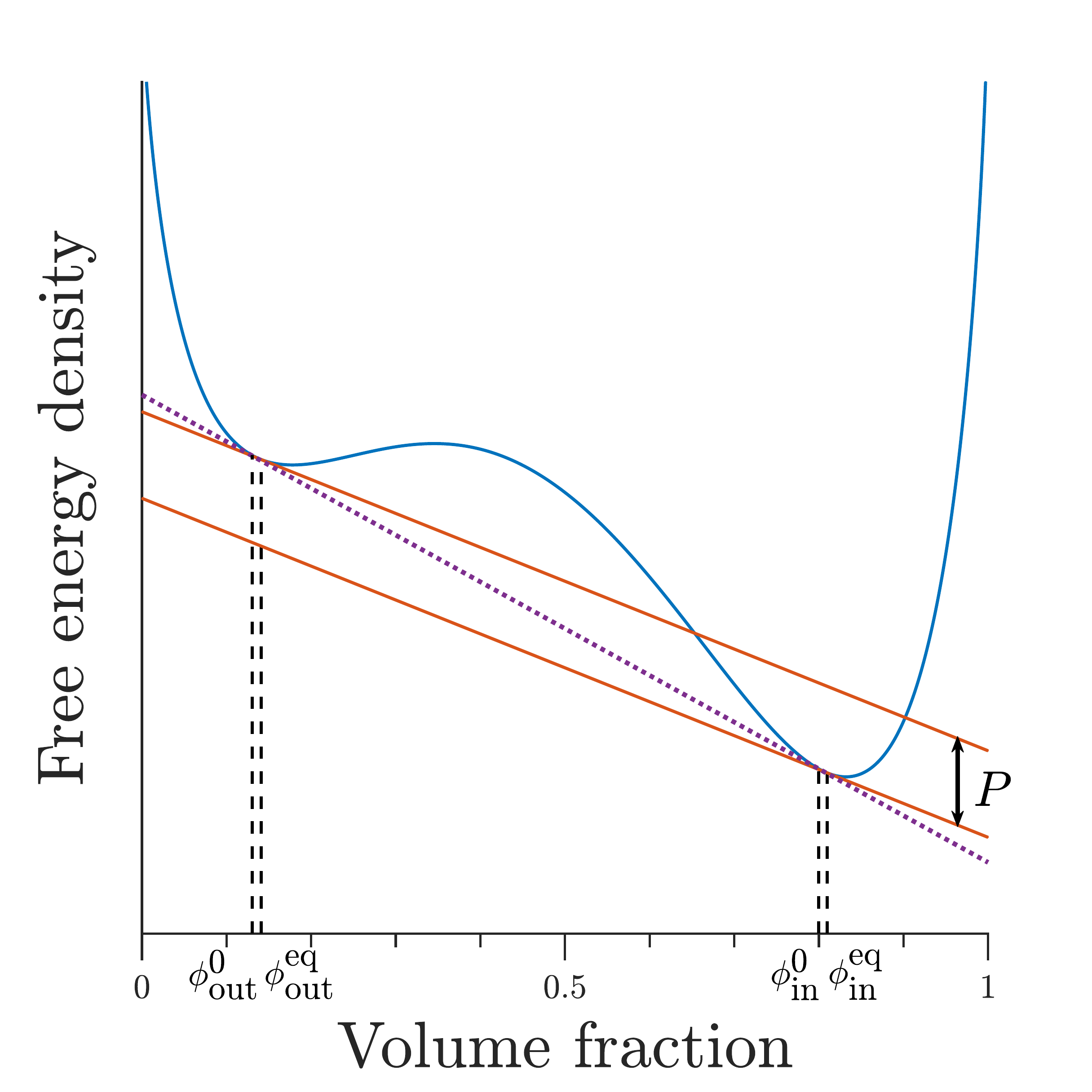}
	\caption{Maxwell construction showing the equilibrium volume fraction of the droplet phase, $\phi_\text{in}^\text{0}$, and dilute phase, $\phi_\text{out}^\text{0}$, in the absence of external pressure, Eq. \eqref{S_Eq_Equilibrium_0_both} (purple dotted line) and how these equilibrium values shift to  $\phi_\text{in}^\text{eq}$ and $\phi_\text{out}^\text{eq}$ in the presence of a pressure difference $P$, Eq \eqref{S_Eq_Equilibrium_with_P} (red continuous lines).}
	\label{S_Figure_Double_Maxwell_construction}
\end{figure}

To understand the system presented in the main manuscript we describe a two-phase system and aim to obtain the change in the equilibrium concentrations when an external pressure is added to the droplet. For a demixed system to be stable, the dilute and the droplet phase need to reach an equilibrium in both chemical potential and osmotic pressure. Assuming an infinite system and given the free energy density of the system $f(V,P,\phi)$, the equilibrium volume fractions in the droplet and dilute phase, $\phi_\mathrm{in}^0$ and $\phi_\mathrm{out}^0$, respectively, fulfil the relations \cite{weber2019physics},
\begin{subequations}
	\label{S_Eq_Equilibrium_0_both}
	\begin{align}
	\label{S_Eq_Equilibrium_0_parta}
	0=&f'(\phi_\text{out}^0)-f'(\phi_\text{in}^0) \hspace{25 pt} \text{and}\\
	0=&f(\phi_\text{in}^0)-f(\phi_\text{out}^0)+(\phi_\text{out}^0-\phi_\text{in}^0)f'(\phi_\text{out}^0),
	\label{S_Eq_Equilibrium_0_partb}
	\end{align}
\end{subequations}

\noindent where the free energy density is differentiated with respect to the volume fraction. Note that for clarity we used here a slightly different notation than in the main manuscript. The expressions $\phi_\mathrm{out}^\mathrm{eq}$, $\phi_\mathrm{out}^\mathrm{0}$, and $\phi_\mathrm{in}^\mathrm{eq}$ here used, correspond to $\phi_\mathrm{eq}$, $\phi_0$, and $\phi_\mathrm{in}$ in the main manuscript, respectively. These equations can be solved using a Maxwell construction as shown in Figure \ref{S_Figure_Double_Maxwell_construction} (purple dotted line). Analogously, when there is a pressure jump $P$ at the interface between phases the new equilibrium concentrations $\phi_\text{in}^\text{eq}$ and $\phi_\text{out}^\text{eq}$ reach a pressure balance,
\begin{equation}
0=f(\phi_\text{in}^\text{eq})-f(\phi_\text{out}^\text{eq})+(\phi_\text{out}^\text{eq}-\phi_\text{in}^\text{eq})f'(\phi_\text{out}^\text{eq})+P.
\label{S_Eq_Equilibrium_with_P}
\end{equation}
The change in equilibrium concentration due to $P$ can be seen in the corresponding Maxwell construction of this system, shown in Figure \ref{S_Figure_Double_Maxwell_construction} (red continuous lines). If the dense phase is tightly packed the molecules can not compress further and thus $f(\phi_\text{in}^\text{eq})\approx f(\phi_\text{in}^{0})$. Using this approximation, Eqs. \eqref{S_Eq_Equilibrium_0_partb} and \eqref{S_Eq_Equilibrium_with_P} can be combined as follows
\begin{multline}
0=f(\phi_\text{out}^\text{0})-f(\phi_\text{out}^\text{eq})-(\phi_\text{out}^\text{0}-\phi_\text{in}^\text{0})f'(\phi_\text{out}^\text{0})
+(\phi_\text{out}^\text{eq}-\phi_\text{in}^\text{eq})f'(\phi_\text{out}^\text{eq})+P.
\label{S_Eq_IntermediateEq}
\end{multline}

Assuming that the pressure increase produces a small change in concentration we can approximate the free energy at the new volume fraction up to first order,
\begin{equation}
f(\phi_\text{out}^\text{eq})\approx f(\phi_\text{out}^\text{0}) +f'(\phi_\text{out}^\text{0})(\phi_\text{out}^\text{eq}-\phi_\text{out}^\text{0}) \; .
\end{equation}

With this approximation and using $\phi_\mathrm{in}^\mathrm{eq}\approx \phi_\mathrm{in}^\mathrm{0}$, Eq. \eqref{S_Eq_IntermediateEq} reduces to
\begin{equation}
0=(\phi_\text{out}^\text{eq}-\phi_\text{in}^\text{eq})\left[f'(\phi_\text{out}^\text{eq})-f'(\phi_\text{out}^\text{0})\right]+P \; .
\end{equation}

Assuming strong phase separation $\phi_\text{in}^\text{eq}\gg\phi_\text{out}^\text{eq}$ and using the definition of chemical potential $\mu=\nu f'(\phi)$, where $\nu$ is the molecular volume of the phase separating material; we obtain an expression for the change of the chemical potential in the dilute phase, due to a change in pressure,
\begin{equation}
P=c_\mathrm{in}^\mathrm{eq}\left[\mu_\text{out}^\text{eq}-\mu_\text{out}^\text{0}\right]
\end{equation}
where $c_\mathrm{in}=\nu\phi_\mathrm{in}$ is the material concentration inside the droplets. Finally, assuming an ideal dilute phase we can approximate $\mu(\phi_\mathrm{out})\approx k_\mathrm{B}T\log{\phi_\mathrm{out}}$, with $k_\mathrm{B}$ the Boltzmann constant and $T$ the absolute temperature, and obtain \cite{rosowski2019elastic}
\begin{equation}
\phi_\mathrm{out}^\mathrm{eq}\approx \phi_\mathrm{out}^\mathrm{0}\exp{\left(\dfrac{P}{c_\mathrm{in}k_\mathrm{B}T}\right)}.
\label{S_eqn:Phi_eq}
\end{equation}
Using the notation of the main manuscript: $\phi_\mathrm{out}^0=\phi_0$ and $\phi_\mathrm{out}^\mathrm{eq}=\phi_\mathrm{eq}$, Eq. \eqref{S_eqn:Phi_eq} is equivalent to Eq. 1 of the main text.

\subsection*{Derivation of the coarse-grained model}
\begin{figure}[h!]
	\centering
	\includegraphics[width=0.75\linewidth]{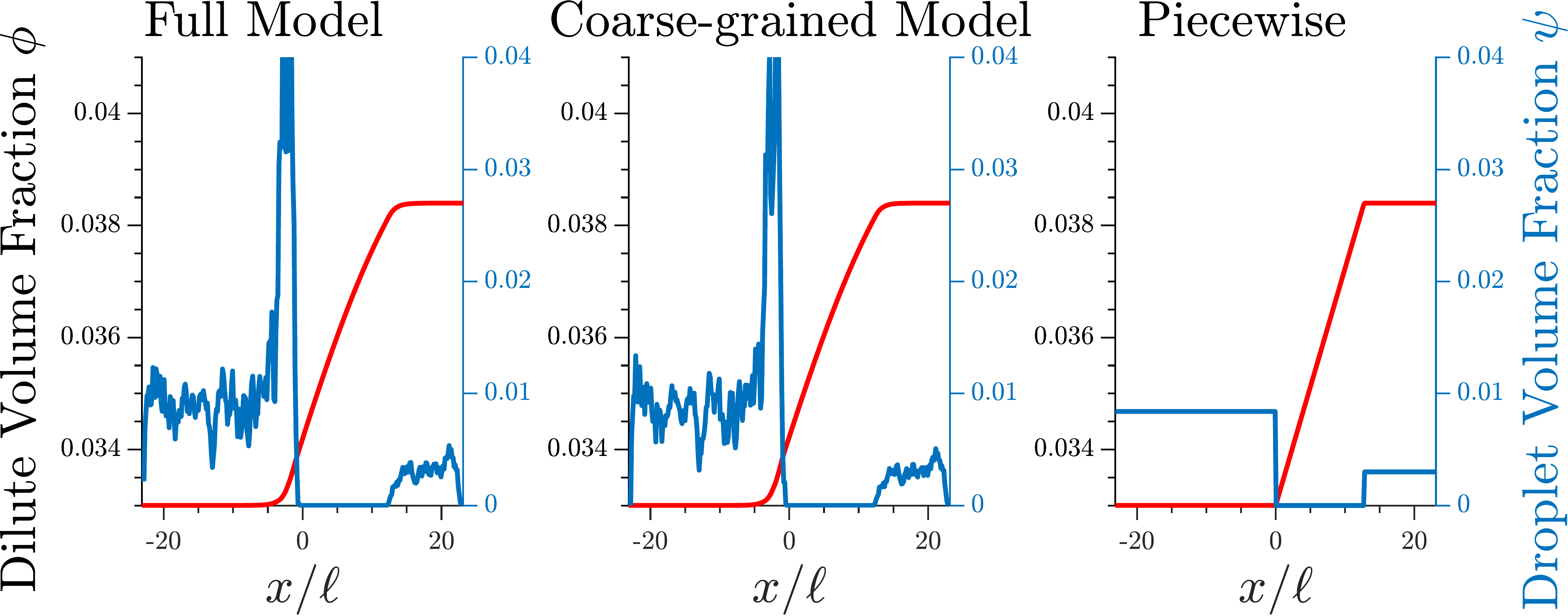}
	\caption{Comparison of the three models used in this work (at a late time point): Full Model Equation (Eqs. 4-5), Reduced Model Equation (Eqs. 8), and Piecewise Model Equation \ref{Eq_Piecewise_model}}
	\label{Figure_triple_comparison}
\end{figure}
In this section we average over individual droplets and deduce an equation for the volume fraction $\psi$ of the material present in the droplet phase, it is defined as
\begin{equation}
\label{S_Eq_Psi_definition}
\psi(\vec{x})=\phi_\text{in}\dfrac{\iiint \sum V_i \delta(\vec{x_i}-\vec{x})dV}{\iiint dV} \, ,
\end{equation}
where the integrals are performed over a small volume V used for coarse graining the system. Assuming monodispersity inside such a volume, the average radius $\langle R \rangle$ of the droplets can be approximated by
\begin{equation}
\label{S_Eq_Mean_R}
\langle R \rangle^3\approx \dfrac{3\psi}{4\pi\phi_\text{in} n}
\end{equation}
where  $n(\vec{x})$ is the droplet number density, defined as the number of droplets present in each discretization volume divided by $V$. If the droplets have similar enough size, then they change volume at similar rates and $n$ does not change over time. From the dynamical equation for the radius of the full model (Eq. (4) in the main manuscript) we obtain the growth rate of a droplet of volume $V_i$
\begin{equation}
\dfrac{dV_i}{dt}=-\dfrac{J_i}{\phi_\mathrm{in}}=\dfrac{4\pi D R_i}{\phi_\text{in}}\left(\phi-\phi^\text{eq}\right) \; .
\end{equation}
Coarse-graining this equation, we get

\begin{equation}
\partial_t\psi=\dfrac{4\pi D \left(\phi-\phi^\text{eq}\right)}{V}\iiint\sum_i R_i\delta(\vec{x}-\vec{x_i})dV,
\end{equation}
where the integral is again performed over $V$. It can be approximated by the average radius multiplied by

\begin{equation}
\iiint\sum_i R_i\delta(\vec{x}-\vec{x_i})dV\approx \langle R\rangle V n(\vec{x})\, , 
\end{equation}

\noindent which in turn can be approximated using Eq. \eqref{S_Eq_Mean_R}, 

\begin{equation}
\partial_t\psi=4\pi n D \left(\phi-\phi^\text{eq}\right)\left(\dfrac{3\psi}{4\pi\phi_\text{in} n}\right)^{1/3} \; .
\end{equation}
The equation for $\phi(\vec{x},t)$ comes directly from applying the definition of $\psi(\vec{x},t)$ to the dynamical equation for $\phi$ of the full model (Eq. (5) of the main manuscript), thus yielding the reduced model 

\begin{subequations}
	\begin{align}
	\partial_t\psi&=    
	D\left(\phi-\phi^\text{eq}\right)\left(\dfrac{48\pi^2n^2\psi}{\phi_\text{in}}\right)^{1/3} \, ,\\
	\partial_t\phi&=D\nabla^2\phi-\partial_t\psi \, ,
	\label{S_Eq_Reduced_Model_partb}
	\end{align}
\end{subequations}
which we use in the main text.

\section*{Derivation of the dissolution front dynamics}
Following the arguments given in the main manuscript, we now calculate the front dynamics at late times, where the front invading the stiff side has travelled a distance $L$ which is much bigger than the transition length, $L\gg w$.\\

In the area devoid of droplets the dynamics reduce to a simple diffusion equation whose boundary conditions are given by $\phi_\mathrm{eq}$ on the soft and stiff sides. If the front moves slow enough we can do a steady state approximation and assume that the area devoid of droplets reaches equilibrium quickly compared to the front speed. For a simple domain we can solve the diffusion equation. In particular, if the elasticity gradient is always in the same direction, we obtain a piecewise approximation dividing the system in three parts along the elasticity gradient direction as follows, 
\begin{subequations}
	\label{Eq_Piecewise_model}
	\begin{align}
	\psi&=\psi_{\text{soft}},&\phi&=\phi_{\text{soft}},&x<0\\
	\psi&=0,&\phi&=\phi_\text{soft} + \frac{x}{L}\Delta\phi,&0<x<L\\
	\psi&=\psi_{\text{stiff}},&\phi&=\phi_{\text{stiff}}, &x>L,
	\end{align}
\end{subequations}
where $L$ is the front position, $\Delta\phi=\phi_\mathrm{stiff}-\phi_\mathrm{soft}$ with  $\phi_\mathrm{stiff/soft}=\phi_\mathrm{eq}(E_\mathrm{stiff/soft})$, $x=0$ is chosen to match the position where droplets start to grow instead of shrink, and $\psi_\text{soft/stiff}$ are the droplet volume fractions after initial growth. A comparison of the volume fractions $\phi$ and $\psi$ between simulations and the piecewise approximation are shown in Figure \ref{Figure_triple_comparison}. Integrating Eq. \eqref{S_Eq_Reduced_Model_partb} over the length of the stiff side yields an equation for the flux at $x=0$,
\begin{equation}
\partial_t\int_0^L\phi dx+\partial_t\int_L^\infty(\phi+\psi)dx=-D\partial_x(\phi+\psi)|_{x=0}\;.
\end{equation}
Using the piecewise approximation, Eqs. \eqref{Eq_Piecewise_model}, the integrals can be easily evaluated, so

\begin{equation*}
\partial_t L \dfrac{\phi_\text{soft}+\phi_\text{stiff}}{2} -\partial_t L (\phi_\text{stiff}+\psi_\text{stiff})=-D\dfrac{\phi_\text{stiff}-\phi_\text{soft}}{L}\;.
\end{equation*}
This equation can be rewritten as
\begin{align}
\partial_t L=\dfrac{\alpha}{2L} && \text{with} && \alpha=4D\left(1+\dfrac{2\psi_\mathrm{stiff}}{\Delta\phi}\right)^{-1}
\;.
\end{align}
Finally, solving for $L$, we find a diffusive motion for the front,
\begin{equation}
\label{Eq_Predicted_alpha}
L=\sqrt{\alpha(t-t_0)},
\end{equation}
where $t_0$ is such that $L(t_0)=0$. This expression shows that the front moves diffusively, with a speed that increases with the elasticity difference and decreases with higher volume fractions of droplet material $\psi$. A comparison between this approximation and the measured front speed is presented in the main manuscript.





\subsection*{Supplementary videos}

\textbf{Supplementary video 1}: Example of a typical simulation with a one dimensional elasticity gradient. Model parameters are $\phi_0=0.033$, $\phi_\mathrm{in}=1$, $\psi_\mathrm{stiff}=0.09\phi_0$, $E_\mathrm{stiff}=0.15\hat{E}$, $E_\mathrm{soft}=10^{-4}\hat{E}$, and $w=1.4\ell$. \\ 

\noindent\textbf{Supplementary video 2}: Different visualization of the elastic ripening process. Video showing the radii and position of  droplets along with the one dimensional average of the dilute phase over time. Same simulation as in Fig. 1 and Supplementary video 1. \\

\noindent\textbf{Supplementary video 3}: Simulation with a two dimensional sinusoidal elasticity profile. Same parameters as in Fig. 5. \\ 

\noindent\textbf{Supplementary video 4}: Simulation with a two dimensional  elasticity profile in the shape of the Max Planck Society's logo.\\ 

\end{document}